# Chiral Magnetism in an Itinerant Helical Magnet, MnSi
# - An Extended $^{29}$Si NMR Study -

Hiroshi Yasuoka*[1,2], Kiyoichiro Motoya[3], Mayukh Majumder**[1], Sebastian Witt[4], Cornelius Krellner[4] and Michael Baenitz[1]

*[1] Max Plank Institute for Chemical Physics of Solids, 01187 Dresden, Germany*
*[2]Advanced Science Research Center, Japan Atomic Energy Agency, Tokai, Ibaraki, 319-1195 Japan*
*[3] Department of Physics, Faculty of Science and Technology, Tokyo University of Science, 2641 Yamazaki, Noda, Chiba, 278-8510, Japan*
*[4]Institute of Physics, Goethe-University Frankfurt, 60438 Frankfurt am Main, Germany*

The microscopic magnetism in the helical, the conical and the ferro-magnetically polarized phases in an itinerant helical magnet, MnSi, has been studied by an extended $^{29}$Si NMR at zero field and under external magnetic fields. The temperature dependence of staggered moment, $M_Q(T)$, determined by the $^{29}$Si NMR frequency, $\nu(T)$, and nuclear relaxation rate, $1/T_1(T)$ is in general accord with the SCR theory for weak itinerant ferromagnetic metals and its extension. The external field dependence of resonance frequency, $\nu(H)$, follows a vector sum of the contributions from atomic hyperfine and macroscopic fields with a field induced moment characteristic to the itinerant magnets. A discontinuous jump of the resonance frequency at the critical field, $H_c$, between the conical and the polarized phases has also been found that suggests a first order like change of the electronic states at $H_c$.

**Introduction:** MnSi is an intermetallic compound with B20 crystal structure, and has been shown experimentally to order magnetically at $T_c$=29.6 K[1)]. The spin structure was predicted first from the analysis of $^{55}$Mn NMR spectra as a helical[2)] and subsequently determined by neutron diffraction to be a long-period (180 Å) in the [111] direction)[3)]. Appling external magnetic field the helical structure changes progressively to the conical and to the ferro-magnetically polarized phase above a critical filed, $H_c$, (0.65 T at 4.2 K). It has also been known that MnSi is a typical example of a weakly itinerant magnet from experimental[4-5)] and theoretical studies[6-7)]. Recently, MnSi has also attracted considerable attention in view of the Skyrmion phase observed in the *M-H* phase diagram near $T_c$ and $H_c$[8-9)]. Although our final goal of the present project is to explore the Skyrmion dynamics by NMR technique, in this letter we deal with the microscopic magnetic properties in the chiral phases obtained by the $^{29}$Si NMR studies in MnSi that has been extended from previous NMR reports[2,4,10)]. We believe that the understanding of spin dynamics in the chiral and the polarized phases is indispensable to attack the Skyrmion physics by the next stage of NMR investigation.

**Experimental:** The $^{29}$Si NMR in MnSi at zero field has been examined in three different samples; zone refined poly-crystalline sample prepared by induction melting[2)] (Sample-1), single crystal grown by the Czochralski method[3)] (Sample-2) and 99.77% $^{29}$Si enriched MnSi (Sample-3). Alt-hough sample-1 and 2 were prepared previously about 40 years ago for NMR and neutron diffraction experiments[2-4], Sapmle-3 was newly prepared for the present experiment using a tube furnace with the sample enclosed in a Nb-crucible. Magnetization measurements were performed with a commercial SQUID VSM from Quantum Design for the single crystal. NMR measurements have been made on Sample-1 for the helical and the conical phases, and Sample-3 for the polarized phase by the standard pulsed spin-echo technique. The reason for the former is due to that it has the sharper spectrum among them, giving accurate determination of the peak frequency and the nuclear spin-lattice relaxation rate, $1/T_1$. In the polarized phase, however, we could only observed signal with $^{29}$Si enriched sample (see below the detail). NMR spectra were taken by the frequency sweep method at zero field or at a constant field. In order to avoid any artificial broadening Fast Fourier Transformation (FFT) signals were summed across the spectrum to produce the frequency spectrum (FFT Sum.). $1/T_1$ was measured by the inversion recovery method as a function of temperature and external field. The recovery of nuclear magnetization after an inversion pulse was a single exponential type and the data were well fitted by $M(\mathrm{t})=M_\mathrm{o}[1\text{-}2\exp(\text{-}\mathrm{t}/T_1)]$ for $^{29}$Si nuclei ($I$=1/2) which has no nuclear quadrupole interaction and is expected to be a single relaxation process.

**Results and discussions:** Before describing $^{29}$Si NMR results, in Fig. 1(a) and (b) we show respectively the temperature dependence of magnetic susceptibility and the external field dependence of magnetization for the single crystal (Sample-2) where the field was applied to the [001] direction. As shown in Fig. 1(a) the magnetic susceptibility follows perfectly the Cure-Weiss law below 300K with $P_\mathrm{eff}$=2.35(±0.01) $\mu_\mathrm{B}$ and $\theta$ =29.87(±0.01) K. These values agree well with reported values[1, 4]. Fig. 1(b) shows the data for magnetization process, $M(H)$, at representative temperatures, showing typical behavior of closing the cone angle in the conical phase to the polarized ferromagnetic phase at $H_\mathrm{c}$ (0.65 T at 5 K). Essentially the same results have been observed in other samples. It should be noted here that the saturation moment extrapolated toward zero field is 0.41(±0.03) $\mu_\mathrm{B}$ at 5K which is about one fifth of the paramagnetic moment (2.35 $\mu_\mathrm{B}$), and $M(H)$ in the polarized phase has a substantial high field susceptibility, $\chi_\mathrm{hf}$=7.67(± 0.03)$\times10^{-7}$ $\mu_\mathrm{B}$/Oe.atom. These facts are clearly showing that the magnetism in MnSi is considered to be an itinerant in its nature[6].

In order to check the quality of samples, the zero-field $^{29}$Si NMR spectra have been examined at 4.2 K. The spectra shown in Fig. 2(a) indicate that all samples have low frequency tail which may be associated with Si atoms having deficiency of Mn atom in their near neighbors. Actually, an EDX analysis of Sample-3 indicates 3.5% Mn deficiency. As is seen in Fig. 2(a) we do not have perfect stoichiometric sample in this experiments. Therefore, Sample-1 was used for NMR in the helical/conical phase and Sample-3 in the polarized phase.

The temperature dependence of center frequency, $\nu(T)$, is shown in Fig. 2(b), where the data were obtained up to 22 K ($T/T_\mathrm{c}$=0.74) by a Gaussian fit using 50% of the intensity data from its maximum value. Because the Sample-1 was not $^{29}$Si enriched and $T_1$ becomes so short at high temperatures (~3 μsec at 20K) that we could not obtain good signal to determine the center frequency with enough accuracy above 22 K. The simple spin-wave type fit, ($\nu(T)$~ $T^{3/2}$), of the data below 8 K yields an extrapolated resonance frequency at 0 K to be ν(0)=19.992(±0.005) MHz which corresponds to the internal field of $H_n(0)$=2.364(±0.006) T at the Si sites. ν($T$) is a good and accurate measure of the temperature dependence of staggered moment, $M_Q(T)$, and predicted theoretically by the extended SCR theory for itinerant helical magnets[11] as,

$$\nu(T) = \frac{{}^{29}A_{hf}}{g\mu_B h} \cdot \langle M_Q \rangle_T = \nu(0)\left\{\frac{(1-t^{4/3})}{1-2/3\cdot t^{4/3}}\right\}, \tag{1}$$

where $t=T/T_c$ with $T_c$=29.87 K for MnSi. The solid curve in Fig. 2(b) shows the calculated result of eq. (1) using the experimental value of ν(0). One can observe an excellent agreement between eq. (1) and the experimental data. Nevertheless, it should be noted that this extended SCR theory is only based on an extension of the weakly ferromagnetic metals using a simplified single Fermi surface to include the spin density fluctuation around small $Q$, (a long period helical structure).

The external field dependence of $^{29}$Si NMR frequency, ν($H$), in the helical, the conical and the polarized ferromagnetic phases has been investigated at 4.2 K and the results are shown in Fig. 3(a). Here, Sample-1 was used for fields below $H_c$ while Sample-3 for above $H_c$ because that the S/N ratio is substantially reduced due to lack of signal enhancement in the polarized phase[2] and it was indispensable to use the $^{29}$Si enriched sample. Generally, the internal field acting on a nuclear site, $H_n$, in magnetically ordered state is sum of the atomic hyperfine field, $H_a=A_{hf}M_Q/g\mu_B h$, the Lorentz field, $H_L=4\pi M_s/3$, the demagnetizing field, $H_D=N\cdot M_s$, and the external field, $H_o$, where $M_Q$ is the atomic moment of Mn, $M_s$ the spontaneous moment along $H_o$ and $N$ the demagnetizing factor ($N=4\pi/3$ for a spherical sample). Taking a field induced moment characteristic to itinerant ferromagnets into account and referring a vector presentation in Fig. 3(b), $H_n$ under a given $H_o$ can be written as,

$$\begin{aligned} H_n^2 &= (-H_a(\theta)+H_L)^2 + (H_o - H_D(\theta))^2 \\ &\quad -2\cdot\left|-H_a(\theta)+H_L\right|\cdot(H_o - H_D(\theta))\cdot\cos\theta \end{aligned}, \tag{2}$$

where $\theta$ is the angle between $H_o$ and $M_Q$ (cone angle) defined as $\cos\theta=H_o/H_c$. The angular dependence of $H_a(\theta)$ and $H_D(\theta)$ can be written as $H_a(\theta)=H_a+A_{hf}\cdot\chi_{hf}\cdot H_o\cdot\cos\theta$ and $H_D(\theta)= H_D\cdot\cos\theta$. Hence, the resonance frequency at the helical ($\theta=\pi/2$), and the polarized ($\theta=0$) phases are simply given by $\nu_h(0)={}^{29}\gamma_n\cdot(H_n-H_L)$ and $\nu_P(H)={}^{29}\gamma_n\cdot|\{H_n+(1+A_{hf}\cdot\chi_{hf}\cdot H_o)\}|$, respectively. Since there is no demagnetizing field in the helical phase all nuclei sense only the Lorenz field as the macroscopic field, then the zero field resonance should appear at lower frequency than that expected from $H_a$ by $H_L$ (635 Oe corresponding to 0.53 MHz for MnSi). This is clearly observed in Fig. 3(a) where $\nu_P(H)$ is extrapolated to zero field by a least squares fit of the data to $\nu_p(H)$ with $\chi_{hf}=7.67\times10^{-5}\mu_B$/Oe.atom. Using the value of moment extrapolated to zero field in the polarized phase, $M_Q(0)=0.41\ \mu_B$, and correcting $H_n$ by $H_L$, the transferred hyperfine coupling constant at the Si nuclei is obtained to be $^{29}A_{hf}=(-H_n+H_L)/M_Q(0)=-5.61$ T/$\mu_B$. This value is to be compared with $^{29}A_{hf}(p)=-5.52$ T/$\mu_B$ obtained from the Knight shift vs. susceptibility plot at the paramagnetic state[4], assuring that there exists no essential change of $^{29}A_{hf}$ between paramagnetic and ordered states. For the conical phase, assuming that 1) the $A_{hf}$ does not depend on temperature and field, 2) the $\chi_{hf}$ is the same as in the polarized phase, and 3) powder sample has spherical shape, meaning $|H_D|=|H_L|=(4\pi/3)\cdot M_s$, the external field dependence of resonance frequency is calculated using eq. (2) and shown by the solid curve in Fig. 3(a). Although we see a good agreement between calculated field dependence and experimental data, there is a discontinuity of $\nu(H)$ between the conical and the helical phases at $H_c$. Actually, the signal intensity measurements at 0.7 T($\sim H_c$) shows a strong rf power dependence with two different types of signal as shown in Fig. 4(a). In Fig. 4(b), the NMR frequency spectra taken by two different rf power levels of -12dB (high power) and -38dB (low power) are shown. These data clearly indicate that there is a crossover region near $H_c$ with a first order like frequency jump. It is also noted that the $T_1$ increases more than one order of magnitude in the polarized phase, meaning a substantial suppression of the magnetic excitations in the polarized phase. The similar phenomenon was also observed by the $^{55}$Mn NMR in MnSi[2)]. This first order like transition in the electronic state at $H_c$ has not been observed by other macroscopic measurements, even in the $M(H)$ shown in Fig. 1(a).

The temperature dependence of $^{29}$Si nuclear magnetic relaxation rate, $1/T_1$, at zero field and the field dependence at 4.2 K have been measured in great detail and the results are shown in Fig. 5(a) and (b), respectively. Although we had almost perfect fit of the $M_Q(T)$ to the extended SCR theory for itinerant helical magnets, unfortunately the dynamical aspects have not been treated properly in this theory and there is no theory which can be relied on. Therefore, we adopted the SCR theory for weak itinerant ferromagnets, keeping the helical state with long pitch (nearly ferromagnetic) in mind. In this case, the SCR theory predicts $1/T_1 \propto T/M_Q(T)^2$ dependence. Using $M_Q$(T) obtained experimentally in Fig. 2(b), $1/T_1(T)$ is well fitted to the function up to $T/T_c$=0.65 with $T_1T=7.96\times10^{-5}$sec.K, as shown by the solid curve in Fig. 5(a). The field dependence of $1/T_1$ is also shown in Fig. 5(b) for the helical, the conical and the polarized phases. Characteristic features of $1/T_1(H)$ in the conical phase ($1/T_1 \sim H^{-2}$) and the polarized phase ($1/T_1 \sim H^{-1}$) and the reduction of $1/T_1$ in the polarized phase by more than one order of magnitude have not been understood yet. The former must be related to the chiral magnetic excitations under magnetic fields. It should be noted that in spite of reasonable agreement between experiments and the SCR theory we should not rely too much on this theory. The dynamical theory based on the realistic multi-band structure is highly expected to interpret all of the data presented here consistently.

**Concluding Remarks:** We have presented extended $^{29}$Si NMR results and explored the chiral magnetism in the prototype itinerant helical magnet, MnSi. The static magnetic properties can be viewed as a consequence of the extended SCR theory for itinerant helical magnet, in spite of the fact that the theory does not consider the symmetry breaking in B20 crystal structure and the multi-band nature in MnSi. The characteristic field dependence of resonance frequency is understood by taking field induced moments into account. The dynamical properties like the nuclear relaxation could not be treated by the SCR theory properly. We need to have more microscopic and dynamical theory to account for the magnetic excitations in the chiral phases. The DMFT approach using the realistic multi-band scheme may be a plausible way, and development of such theory is expected.

**Acknowledgment:** We thank Prof. K. Ueda, Dr. Y. Yanagi and Dr. U. Roessler for useful discussions from view points of the theory of itinerant chiral magnetism. KM acknowledges the support from a Grant-in-Aid for Scientific Research from the Ministry of Education, Culture, Sports, Science and Technology (No. 24540351)

---

*yasuoka-h@nifty.com

** Present address, Department of Physics, The University of Tokyo, Hongo7-3-1, Bunkyo-ku, Tokyo, 113-8654 Japan.

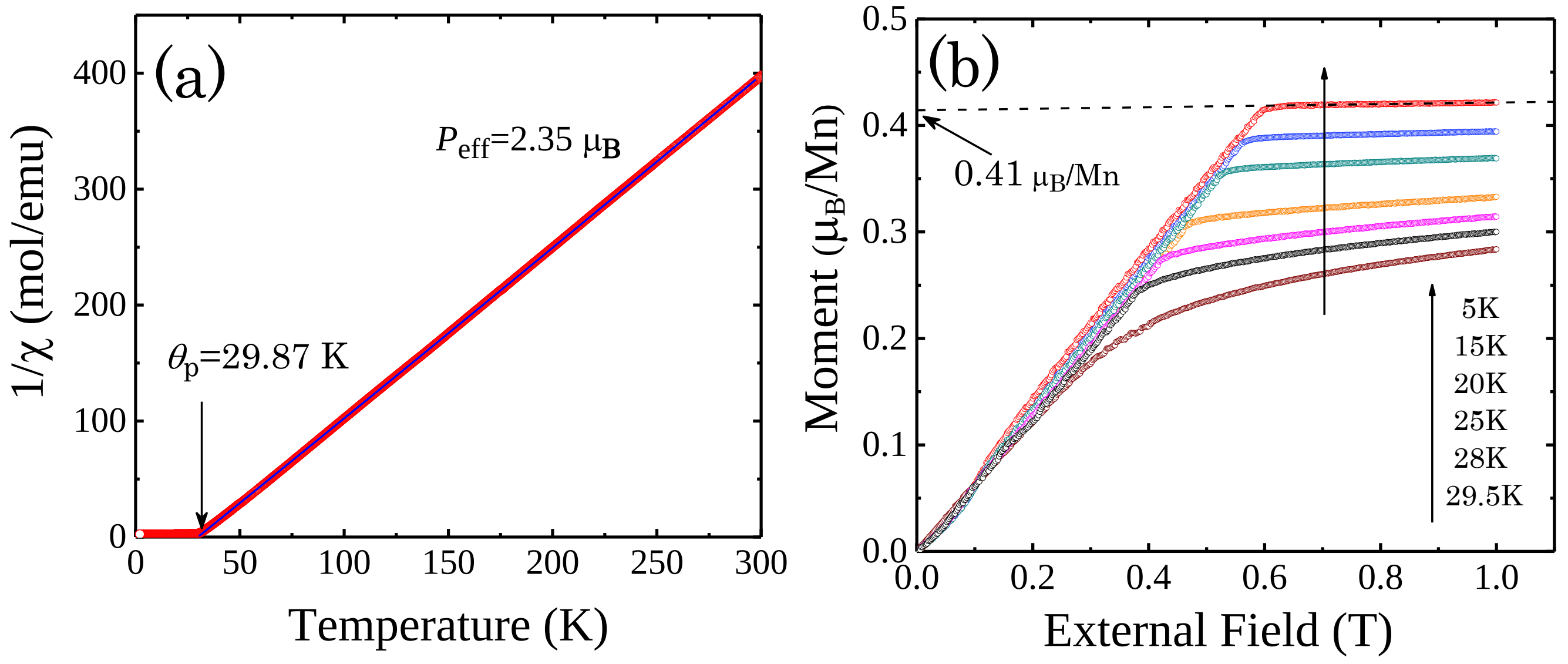

**Fig. 1.** (Color online) (a) Temperature dependence of inverse of the magnetic susceptibility at 0.12 T in MnSi. The Cure-Weiss fit yields $P_{eff}$=2.35(±0.01) $\mu_B$ and $\theta_p$ =29.87(±0.01) K. (b) External field dependence of moment from 5.0 K to 29.5 K in single crystal of MnSi. The external field was applied along [001] direction. From a linear fit of data at 5K between 0.7 T and 1.0 T the saturation moment and high-field susceptibility have been obtained to be 0.41(±0.03) $\mu_B$ and 7.67(±0.03)$\times 10^{-7}$ $\mu_B$ /Oe.atom, respectively.

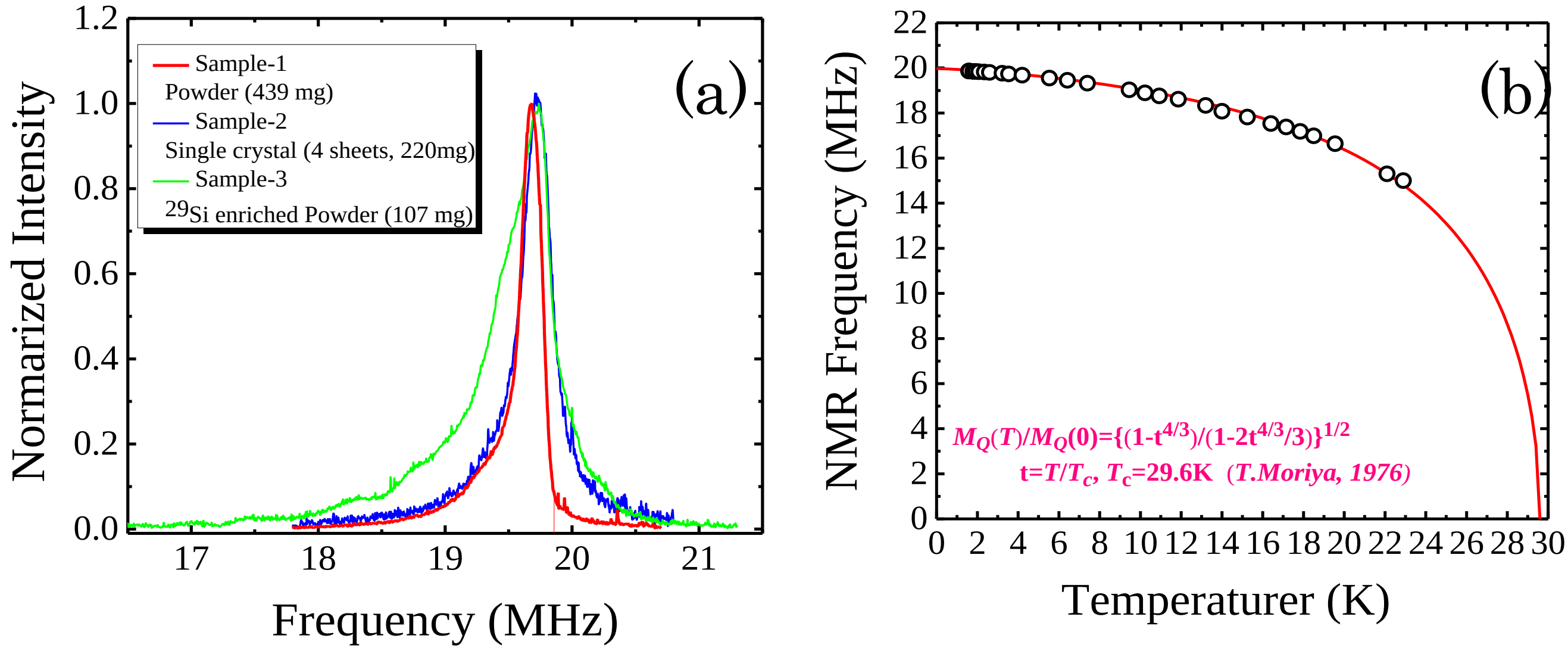

**Fig. 2** (Color online) (a) $^{29}$Si Zero-field NMR spectrum in $^{29}$Si natural abundant (Sample-1, red), 99.77 % $^{29}$Si enriched powder (Sample-3, green) and single crystal (Sample-2, blue) of MnSi. All spectra were taken by the FFT-sum method with 50 kHz step at 4.2 K. (b) Temperature dependence of $^{29}$Si zero-field NMR frequency in MnSi. The solid red line is the temperature dependence expected from the extended SCR theory for itinerant helical magnet (ref. 11).

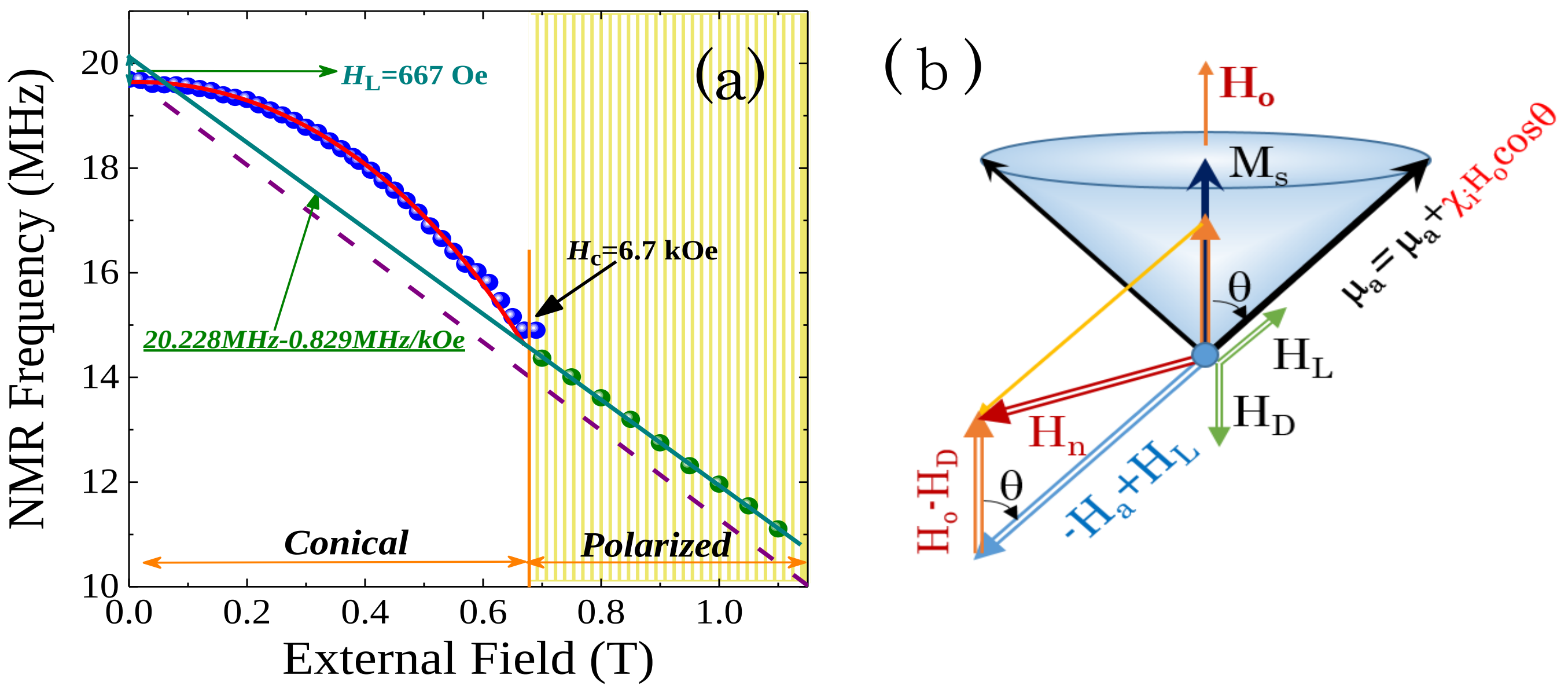

**Fig. 3.** (Color online) (a) External field dependence of $^{29}$Si NMR frequency, $\nu(H)$, at 4.2 K in MnSi. The red curve shows the expected $\nu(H)$ from eq. (1) and green line shows a linear fit of the data for $H>H_c$. Note that the experimental slope (8.29(± 0.02) MHz/T) is slightly smaller than that expected from the nuclear gyromagnetic ratio of $^{29}$Si (8.4578 MHz/T, shown by the broken line) which is attributed to the presence of a field induced moment. (b) An illustration of vector sum for the internal field from various contributions (see text).

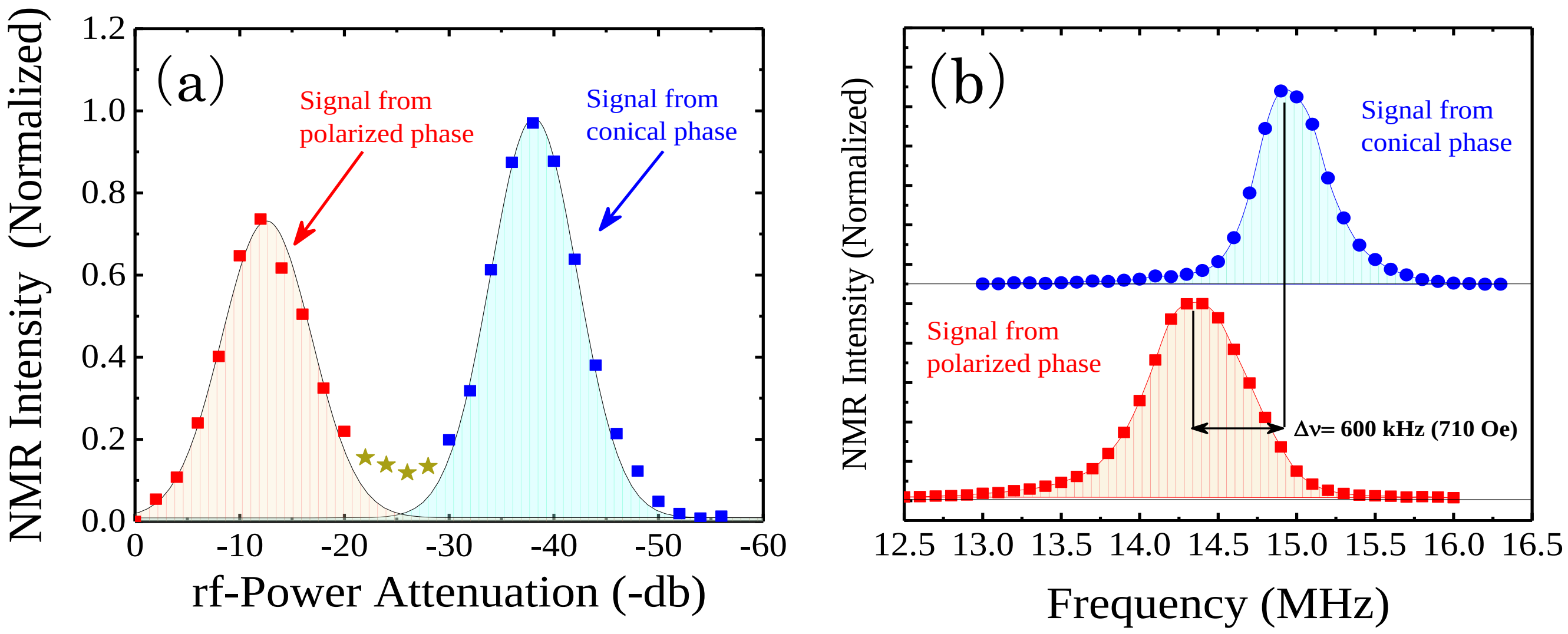

**Fig. 4** (Color online) (a) $^{29}$Si NMR intensity as a function of rf power measured at 4.2 K and under 0.68T with a constant frequency of 14.8 MHz. (b) Associated NMR spectra for high (-12 db) and low (-38 db) power setting, showing coexisting of two types signals with slightly different frequency (~600 kHz).

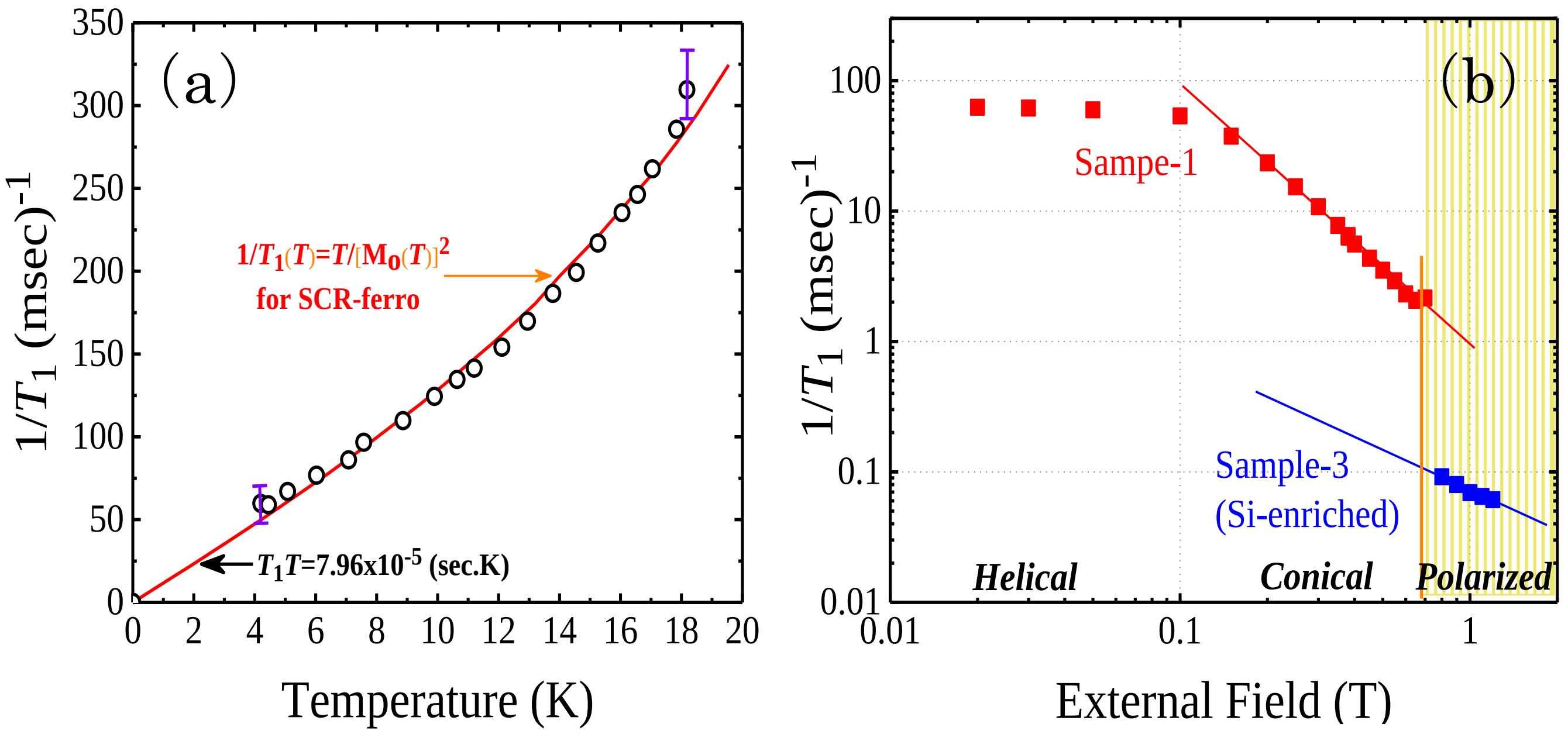

**Fig. 5.** (Color online) (a) Temperature dependence of $1/T_1$ at zero field. A fit of the data to the SCR theory for weak itinerant ferromagnets with experimental $M_Q(T)$ is shown by the red curve. (b) External field dependence of $1/T_1$ at 4.2 K in MnSi.